\documentclass[useAMS,usenatbib]{mn2e}
\usepackage{graphicx}
\usepackage{txfonts} 
\UseRawInputEncoding

\newcommand{\aap}{A\&A}
\newcommand{\aapr}{A\&A Rev.}
\newcommand{\mnras}{MNRAS}
\newcommand{\apj}{ApJ}
\newcommand{\aj}{AJ}

\newcommand{\apjs}{ApJS}
\newcommand{\araa}{ARA\&A}
\newcommand{\nat}{Nature}
\newcommand{\msait}{Mem. S.A.It.}
\newcommand{\pr}{Phys. Rep.}

\catcode`\@=11
\def\gsim{\ifmmode{\mathrel{\mathpalette\@versim>}}
    \else{$\mathrel{\mathpalette\@versim>}$}\fi}
\def\lsim{\ifmmode{\mathrel{\mathpalette\@versim<}}
    \else{$\mathrel{\mathpalette\@versim<}$}\fi}
\def\@versim#1#2{\lower 2.9truept \vbox{\baselineskip 0pt \lineskip
    0.5truept \ialign{$\m@th#1\hfil##\hfil$\crcr#2\crcr\sim\crcr}}}
\catcode`\@=12
\def\msun{\hbox{$M_\odot$}}

\def\yr-1{\hbox{${\rm yr}^{-1}$}}

\def\rsun{\hbox{$R_\odot$}}

\def\t9{\hbox{$t_9$}}

\def\m*{\hbox{$M_{\rm stars}$}}

\def\ho{\hbox{$H_\circ$}}
\def\h50{\hbox{$\ho /50$}}

\begin{document}

\title{On the Ubiquity of Extreme  Baryon Concentrations in the Early Universe}
\author[Alvio Renzini]{Alvio Renzini$^{1}$\thanks{E-mail: alvio.renzini@inaf.it}\\ 
$^{1}$INAF - Osservatorio
Astronomico di Padova, Vicolo dell'Osservatorio 5, I-35122 Padova,
Italy}

\date{Accepted 2024 October 29. Received 2024 October 9. In original form 2024 September 5.}


\maketitle
                                                            

\begin{abstract}
Early JWST observations have revealed the ubiquitous presence in the early Universe, up to $z\sim\! 16$, of extreme baryon concentrations, namely forming globular clusters, extremely dense galaxies that 
may or may not be UV bright, and supermassive black holes in relatively low-mass galaxies. This paper is trying to pinpoint which physical conditions may have favored the formation of such concentrations, that appear to be very common at high redshifts while their formation being progressively more and more rare at lower redshifts. Building on local globular cluster evidence, it is argued that such conditions can consist in a combination of a $\sim\! 10$ Myr extended feedback free time, coupled to low angular-momentum densities  in deep local minima of the ISM vorticity field, where baryon concentrations are more likely to form. It is argued that the former condition would follow from more massive stars failing to explode as supernovae, and the latter one  from low vorticity prevailing in the early Universe, in contrast to later times with their secular increase of the angular momentum density due to the cumulative effect of tidal interactions.

\end{abstract}


\maketitle

\section{Introduction}
\label{sec:intro}
After two years of JWST operations the perception is spreading that we are close to be able to see virtually all existing galaxies within our accessible Universe, in principle becoming able to
 inventory most, if not all the $\sim 300$ billion galaxies within it. The full observational mapping of galaxy evolution is becoming possible, from their first seeds to the present. This has been a very exciting time, with many, unexpected, discoveries (e.g., \citealt{adamo24a}, for an early review). First, the excess of UV-bright galaxies at $z\simeq 9-16$ over the expectation from pre-JWST  theoretical 
models calibrated at lower redshifts (e.g., \citealt{finkelstein23, finkelstein24}). Then the ubiquitous presence of  AGNs with super-massive black holes (SMBH) in relatively low mass galaxies at $z\gsim 4$ (e.g., \citealt{maiolino23a, pacucci23}), reaching extremely high gas densities \citep{maiolino24b}. Furthermore, very compact galaxies ($R_{\rm e}\lsim 200$ pc)  including the so-called {\it little red dots} (LRD, e.g., \citealt{matthee24, pablito24}) with extremely high central densities \citep{baggen24}, comparable to (or exceeding) those of nuclear star clusters in the local Universe \citep{neumayer20}.  Finally, lensed dwarf galaxies at $z\simeq 6-10$ host strings of compact bound clusters \citep{vanzella23, adamo24b} with sizes and masses comparable to those of their likely descendants, i.e., globular clusters such as those in our local Universe.

There remains some ambiguity as to whether the AGNs at high redshifts are real and their estimated  SMBH masses are correct. These are derived from the locally calibrated relation with overall luminosity and (broad line region) Balmer line width \citep{maiolino23a}. On the other hand, an X-ray stack of candidate AGNs shows no appreciable emission \citep{maiolino23b}, well below the level common  among local AGNs with otherwise similar parameters.
Maiolino and collaborators argue that these high-$z$ AGNs could be Compton thick, hence with no X-rays escaping from them. Alternatively, there may be no AGN at all, and the line broadening may be just due to the extreme compactness of these early galaxies (e.g., \citealt{akins24, baggen24}). Moreover, LRD galaxy-AGN decomposition indicates that the host galaxy itself is extremely compact \citep{akins24}.Thus, no matter whether AGNs dominate or not in these galaxies, the extreme baryon concentrations  remain, in one form or another. If no AGNs, then the estimated stellar masses and densities would be even higher \citep{baggen24}, if AGN dominate, so would SMBHs, the most extreme baryon concentrations, several per cents the stellar mass of the host \citep{maiolino23a, pacucci23,  inayoshi24}. 

In brief, the young Universe, few hundred million years after the Big Bang, was extremely prolific in producing the most extreme baryon concentrations we know of, such as globular clusters, nuclear clusters and extremely compact galaxies, and supermassive black holes. Apparently, not much was contrasting a tendency of
 baryons to cool, sink and reach extreme densities while deepening their own potential wells. Why was it so? Up to now, much of the ongoing discussion about first JWST results has focused on the abundance of these objects, such as their UV luminosity function, mass function and number density, i.e., their global properties, while  relatively less attention has been devoted on the physical conditions that may have favoured the formation of such extreme baryon concentrations.

 \begin{figure*}
\includegraphics[height=.4\textwidth,trim={0.6cm 3.2cm 0.6cm 7.5cm},clip]{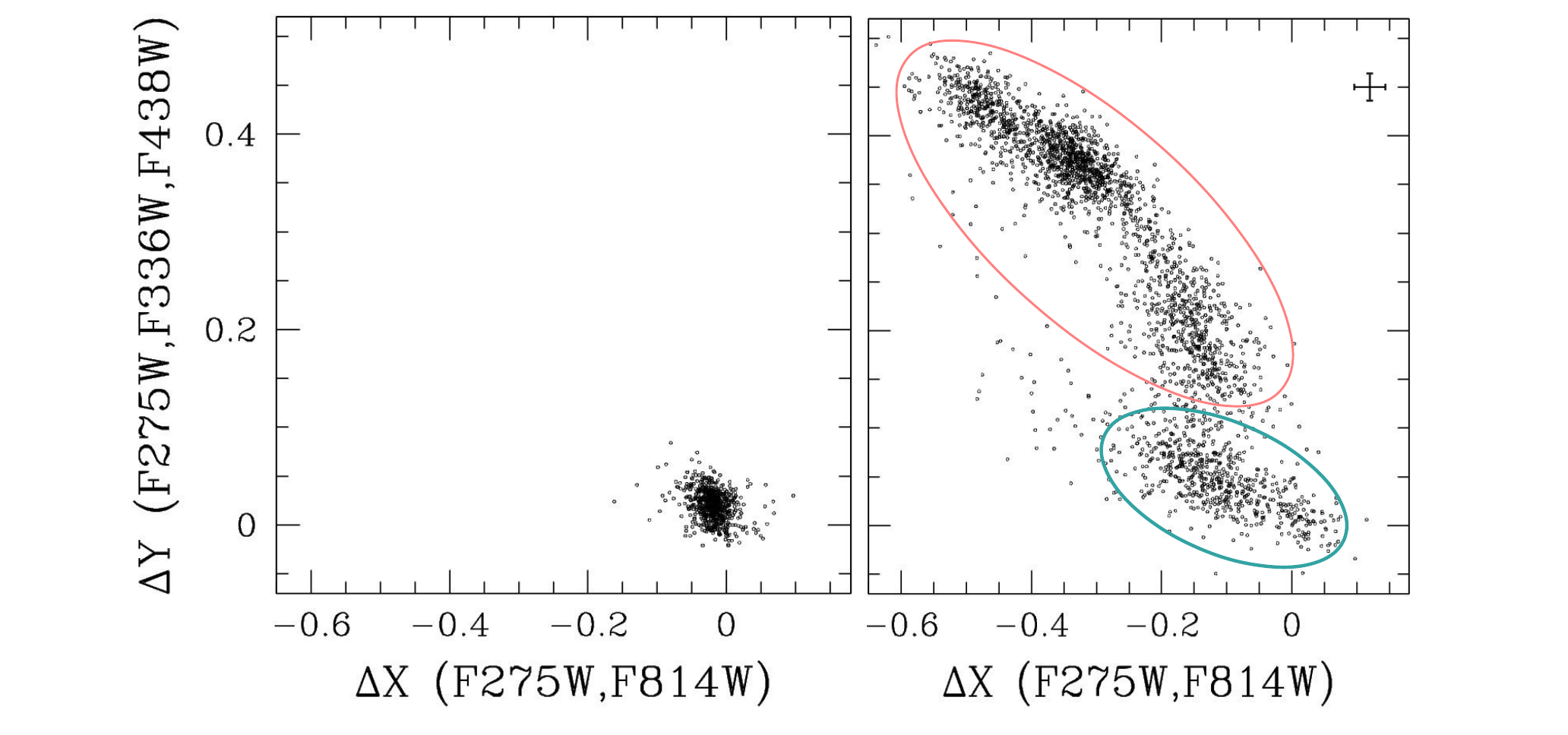}
\caption{Left panel: the chromosome map of a hypothetical chemically homogeneous simple stellar population, where the scatter is only due to photometric errors, compared to (right panel)  the actual chromosome map of the globular cluster NGC 2808 from Milone et al. (2017). Courtesy of Antonino Milone. On the two axis are colour/pseudo-colour combinations of the indicated HST filters. Namely, on the horizontal axis is the star-to-star relative difference of the F275W--F814W colour, and on the vertical axis is the  relative difference of the (F275--F336W)--(F336--F438W) pseudo-colour.}
\label{fig:fig1}
\end{figure*}

 This paper explores the possibility to {\it unify} the problems of the formations of such baryon concentrations, looking whether common physical processes may exist that generically 
 favour unimpeded baryonic cooling, collapse, star formation, and more. It is not meant to work out issues to their quantitative details, but rather pinpoint possible connections and laid down 
a scenario that could be further explored, checked and possibly perfected  with contributions by many.
 Globular clusters, the less extreme case of the three, will  perform a major role in this attempt, as they may provide a key to open also the other two doors. This will expand over previous elaborations (\citealt{renzini22, renzini23}, hereafter Paper I and Paper II, respectively).
 
\section{Attempts at understanding what we see}
A number of possible explanations for the excess of UV-bright, massive galaxies at $z=9-16$ over pre-JWST simulations and semi-analytic models have been proposed, focusing on the baryonic physics assumptions that may have been invalid at such high redshifts. Compared to previous assumptions, these include a more top-heavy IMF \citep{harikane23}, stronger burstiness in early galaxies \citep{shen23}, stronger evolution of dust attenuation \citep{ferrara23} and reduced stellar feedback (starting with \citealt{finkelstein23}, and then many others). A top-heavy IMF may help models to produce more UV-bright galaxies, but would not favor baryonic concentrations, actually, it would increase feedback.
Burstiness, per se, would not affect  baryonic concentrations either, in case being the result of them rather their cause. Similarly, dust evolution via expulsion in winds would act contrary to baryon concentrations, rather than favouring them. Among these options,  only a somehow reduced feedback has potentially the capability to facilitate and promote baryon concentrations. So, this paper is focused on it.

\subsection{Insights from Globular Clusters}
The formation of the less extreme of the three kinds of baryon concentrations, globular clusters (GC), may be easier to understand because we see them in formation at high redshift and have abundant information on their progeny in the local Universe, especially on those in our Milky Way. These objects have comparable masses and sizes even at the two extremes of cosmic time, suggesting that they  experienced relatively minor mass/size evolution in the intervening $\sim\! 13$ Gyr.  

It is known since several decades that Galactic GCs are not chemically homogeneous, but host multiple stellar generations with various degrees of carbon and oxygen depletion anticorrelated to nitrogen and sodium enhancements, resulting from proton-capture precesses at high temperatures (see e.g., \citealt{gratton12} and \citealt{bastian18}, for extensive reviews). The ubiquity and varieties of the multiple generation phenomenon has been most extensively documented thanks to an HST Treasury Program targeting 57 GCs \citep{piotto15}. This study has used a combination of UV and optical passbands sampling molecular band blanketing by OH, NH,
CN and CH, thus allowing the distinction of stars with various degrees of CNO nuclear processing. A particularly effective  combination of colours and pseudo-colours, then dubbed {\it chromosome map}
(ChM) was introduced in \cite{milone17} and there displayed for all the 57 GCs. On top of a common basic pattern, distinguishing first (1G) and second generation (2G) stars, large  differences exist from one GC to another and, most importantly, 2G stars are always the majority, ranging from $\sim\! 50$ to over $\sim\! 80$ per cent of the total population of each cluster.

As an example, Figure 1 shows the ChM of the globular cluster NGC 2808 (right panel), compared to the expected ChM for a simple stellar population with the same photometric errors. 
Clearly, GCs are far too complex objects than SSPs. Two main groups of stars can be distinguished: the 1G stars (encircled green) characterised by CNO proportions analogue to those of field stars of the same metallicity [Fe/H], and the 2G ones (encircled red) with altered CNO  proportions. This sequence is one of increasing nitrogen and helium and decreasing carbon and oxygen. Notice that 2G stars indeed largely outnumber the 1G ones. Our reading of this ChM is the following (cf. Paper I). A first burst of star formation generated the 1G, then followed by a sequence of bursts while the composition of the ISM was changing. This pattern is common to all studied Galactic GCs \citep{milone17}: the first burst did not clear the nascent cluster from its ISM nor it hampered the occurrence of the 2G bursts. {\bf There was no negative feedback between one burst and the following one(s).} Actually, more contaminated gas had to be flowing into the nascent cluster to sustain the formation of 2G stars.

Moreover, in most cases 2G stars exhibit only products of proton-capture processes in a high-temperature, hydrogen-rich environment (e.g., \citealt{gratton12}), but {\bf are not enriched by supernova products}. Thus, there was no supernova feedback and no contamination by heavy elements from supernovae during the formation of both 1G and 2G stars, i.e., during the whole cluster formation.  The simplest interpretation of this evidence is that there were no supernovae (Paper I). Once 1G stars formed, an unimpeded runaway, an overcooling catastrophe followed, leading to very high efficient stars formation, i.e., a very high fraction of the available baryons being turned into stars. Yet, this did not happen in a very short time, such as, say, $\sim\! 1$ Myr, because processing stellar envelopes through p-captures and delivering the products takes time: a stellar evolution timescale, i.e., the main sequence lifetime of the stars producing the processed materials.  This requires an extended {\it feedback free time} (FFT), such that stars 
in a substantial range of masses had time to contribute their CNO-processed products, and so pile up enough row materials for manufacturing the second stellar generations of globular clusters.

Our {\it ansatz} in Paper I was that massive stars above a critical mass ($\gsim 20\,\msun$, hence with lifetimes of $\sim\! 10$ Myr, or shorter) fail to explode as supernovae, but rather sink silently into black holes without much energy or heavy elements being delivered. This notion is widely entertained on the theoretical side (e.g., \citealt{sukhbold16, limongi18}). For example, these latter authors recommend to use only wind nucleosynthesis above $25\,\msun$, as more massive stars would not end up with a supernova display, an assumption adopted also by \cite{watanabe24}. In this scenario, following an event of star formation, supernova feedback  is delayed by some $\sim\! 10$ Myr, as only after this time interval stars less massive than the $\sim\! 20\,\msun$ limit start to undergo their supernova explosion, thus interrupting further star formation.

Perhaps the stronger argument in favour of  this scenario comes from observations. \cite{smartt15} searched for precursors of supernovae exploded within 30 Mpc from us  and found 45 of them, 
 all less massive than $18\,\msun$,  whereas $\sim\! 13$ above this limit would have been expected for a Salpeter-like IMF slope. This provides some support to our assumption that stars more massive that $\sim\! 20\,\msun$ do not produce an energetic supernova. More recently, \cite{schuyler23} have identified six more core-collapse supernova progenitors, all complying with Smartt's upper mass limit.

To cope with the excess of UV-bright galaxies at $z=9-16$ as revealed by the early JWST observations, \cite{dekel23} have proposed a specific mechanism to reduce stellar feedback at high redshifts. They assume that all massive stars undergo a supernova explosion as a result of their final core collapse, but prior to such events they would produce only weak feedback. This is so because at the low metallicities dominating in the early Universe, stellar winds carry only little kinetic energy, so to ensure an FFT following a burst of star formation, lasting of the order of $\sim\! 1$  Myr. As argued in Paper II, this may well alleviate the tension between pre-JWST models and the mentioned JWST results, but would fall short of accounting for the multiple generations in GCs. This is so for two reasons: 1) A  $\sim\! 1$ Myr FFT is too short to allow sufficient production of nuclearly-processed material for the formation of 2G stars, and 2) multiple stellar generations are not limited to low-metallicity GCs, but are present at all metallicities, up to solar, such as in the case of the bulge GCs NGC 6528 and NGC 6553 \citep{kader22}, the two most metal-rich GCs in the Galaxy. Thus, to account for the multiple generation in GCs, a much longer FFT (of the order of $\sim\! 10$  Myr) is required, at all metallicities up to solar. In any event, very high gas densities promote strong dissipation and  tend to reduce the effect of stellar wind and radiation feedback \citep{elmegreen17, dekel23}, especially if a runaway converging flow keeps pressurising the forming cluster.

Globular clusters do not form out of the blue, but inside host galaxies. The metal-poor ones inside dwarfs, the metal-rich ones inside more mature galaxies. Actually, only a host galaxy can provide enough material for the formation of second generation stars, and do so as a converging flow feeding the nascent cluster (cf. Paper I). The 30 Doradus star-bursting region in LMC may provide a local analog to this situation, where the central R136 cluster (a protoglobular?) is surrounded by an extended  ($\sim\!\!$ 100 pc)  star-forming region \citep{schneider18}. As mentioned by \cite{farjon24}, it would be most interesting to ascertain whether the gas is still converging towards R136,  thus sustaining further star formation, or it is being blown away by feedback effects. An intriguing aspect on the connection between protoglobular clusters and host galaxies is that  the quintet of clusters at $z\sim\! 10$ appears to overshine the host  galaxy  \citep{adamo24b}.
 
 
\subsection{Is low feedback sufficient to account for extreme baryon concentrations?}
If an extended FFT is needed at all metallicities, so why this ubiquity of extreme baryon concentration is so prominent only in the early Universe? If an extended FFT is a generic property of star formation, due to the temporary lack of supernova feedback, why its effects are not equally apparent at all redshifts? May be because a long FFT is not a sufficient condition to lead to extreme baryon concentrations. Some other condition may be necessary for their occurrence. This additional condition could be low angular momentum.

Proper motion and radial velocity studies have revealed that GCs do rotate, contrary to earlier expectations  (e.g., \citealt{bellini17,bianchini18} and \citealt{petralia24}) . Yet, they do so with relatively
low velocity, typically just a few km s$^{-1}$ \citep{petralia24}. Such slow rotation in compact objects implies GCs to have low angular momentum, reflecting a low angular momentum of the gas out of which they formed. Compact galaxies at high redshifts, such as the ``little red dots",  also  have low angular momentum, given their small size, even if they rotate relatively fast as in the case of GN-z11 \citep{xu24}. On the other hand, baryons may collapse to small sizes and extremely high densities only if they have low angular momentum to start with.

Following \cite{peebles69}, it is generally accepted that the angular momentum of galaxies is acquired by tidal interactions. At high redshifts, baryons started with low angular momentum (low vorticity) and kept increasing their angular momentum density and vorticity all the way to the present, as the effects of (stochastic) tidal interactions cumulate over time. Thus, the early Universe, besides being denser to start with, was also characterised by minimal vorticity, thus favouring  the collapse of baryons to form globular clusters and compact galaxies. Later on, the capacity to form such compact objects was confined to the deepest minima in the vorticity field, that however became rarer and rarer with time, due to the continuing stirring effect of tidal interactions. A natural outcome of this trend is that the specific frequency of GCs (their number over the field star content at the same metallicity) should decrease with increasing metallicity, as indeed observed (e.g., \citealt{harris06}, and references therein), for both metallicity and vorticity increase with time,  though they do so for different reasons. In summary, a delayed feedback coupled with low vorticity may be responsible for promoting extreme baryon concentrations, more so at high redshifts, but, even if more rarely, also at lower redshifts. One such example may be offered by the compact {\it blue monster} at $z=3.6$ \citep{marques24}. Compact, young massive clusters occasionally form also in the near Universe, especially in mergers (e.g., the Antennae, \citealt{zhang99}) broadening the vorticity distribution, hence generating new minima.

\section{Discussion} 
In this paper it has been argued that, among the proposed solutions  to account for the unexpected frequency of UV-bright galaxies at $z=9-16$, only a temporary reduced stellar feedback has the capacity to favor the formation of extreme baryon concentrations at these redshifts. Such concentrations include forming globular clusters, very compact galaxies such as the ``little red dots" and others, and even oversized SMBHs.

Here, expanding from Paper I and Paper II, it has been argued that an extended 
feedback-free time (FFT) is required to account for the formation of multiple stellar generations in Galactic globular clusters, coupled to second generations having escaped contamination by supernova products. The simplest way to accommodate both these requirements is to assume that most massive stars above a limit (around $\sim\! 20\,\msun$) fail to explode as supernovae but directly collapse to black holes. In this scenario, an FFT of $\sim\! 10$ Myr allows massive stars (most being interacting binaries) to deliver proton-capture processed materials for the formation of second generation stars. 

This critical assumption  (no supernovae during the first $\sim\! 10$ Myr after a burst of star formation) has some empirical support from the identification of supernova precursors \citep{smartt15}, but certainly a more extended sample of such precursors would be required to fully confirm whether stars more massive than $\sim\! 20\,\msun$ 
fail to explode. High-resolution imaging over very wide areas is required for such a test, something that is being provided by the Euclid mission planned to cover 1/3 of the whole sky.  Wide-area surveys from the ground (e.g., LSS/Rubin) could also contribute. Thus, a fairly high fraction of  future supernovae within 30-100 Mpc will have an identified precursor, and the ultimate fate of massive stars is likely to be settled in the near future. And so will be the duration of the FFT past a burst of star formation, with all its consequences for the emergence  of the extreme baryon concentrations at high redshift. This fascinating possible connection between the outcome of core collapse of massive stars and the early Universe, is worth to be further explored.

On the other hand, a link between globular clusters and high-redshift galaxies was soon recognised from the early JWST results, starting with the famous case of the $z=10.6$ galaxy GN-z11 \citep{bunker23}, where the ISM is highly enriched in nitrogen, just as in the case of GC second generation stars (see also \citealt{senchyna24, belokurov23, watanabe24}). So, objects such a GN-z11 may be forming the second generations of their nascent globular clusters. However, also GC first generations stars had to form, and therefore the expectation is that, along with nitrogen-rich galaxies, also  galaxies with {\it normal} nitrogen (i.e., 1G like) should be found at the same redshifts.
  
In Paper I the notion was endorsed according to which massive binary stars are the producers of the material for second generations, yet, this is still to be proven. Also other candidate producers have been entertained over the past decades, with strong partisanships around each option (see \citealt{bastian18}, for a recent review). It is plausible that massive interacting binaries deliver the required proton-capture processed material in the right amount, in the right proportions and at the right time, but there is no guarantee they actually do so. From the theoretical point of view, these issues depend on the recipes adopted for deep mixing of non-convective origin and on the relative timing between nuclear processing and delivery by mass loss: all processes for which a fundamental theory does not exist. By the way, this basic uncertainty is common to all candidate producers. Vindication of the massive binary option could come from direct spectroscopic observations of massive binaries, especially if caught during their common-envelope events. Are they enhanced in nitrogen and depleted in carbon and oxygen?

The recent discovery of an intermediate mass black hole at the centre of the globular cluster $\omega$ Centauri, with a firm lower limit on its mass of  $\sim\! 8,200\,\msun$ \citep{haberle24}, has revitalized the interest on the possible connection between globular clusters and the formation of SMBHs. Indeed, the alleged ubiquity of SMBHs in the early Universe, and the short time available  to grow them, may favor SMBH seeds from direct collapse of supermassive stars (SMS, e.g., \citealt{pacucci23,jeon24}), as opposed to seeds resulting from the merging of many 
stellar-mass black holes. 

SMSs, say, $10^4-10^6\,\msun$, subject to general-relativistic instability, that may be precursors to SMBHs,  have been considered since a long time \citep{fowler66} and have been entertained also in connection with the multiple generations in globular clusters
\citep{denissenkov14}. We do not know whether the black hole at the centre of $\omega$ Cen is the remnant of a SMS, but certainly GCs in formation at high redshifts are the object that may host SMSs, if they exist. Incidentally, they would represent yet another kind of extreme baryon concentration. According to models, SMSs are expected to be rather compact objects, with effective temperatures of 50,000 to 100,000 K \citep{woods20}, hence with a very steep UV slope and could be difficult to distinguish from very young starbursts. Identifying likely candidates would provide targets for the ELT spectroscopic confirmation.

The scope of this paper is to attract attention on the possible links among a variety of distinct astrophysical problems and phenomena, namely the fate of massive stars, the formation of globular clusters, of the compact and/or UV bright galaxies in the epoch of re-ionization and of the formation of intermediate mass and supermassive black holes.
Understanding how globular clusters formed, along with their multiple stellar generations, will help to understand also how the other kinds of extreme baryon concentrations came into being in the early Universe.

\section*{Acknowledgments} 
I wish to thank Antonino Milone for his permission to reproduce here Figure 1, and together with Anna Marino for having involved me in their insightful photometric and spectroscopic research on globular clusters. I warmly  acknowledge Mauro Giavalisco and Eros Vanzella for their constructive comments and encouragement. 
I wish also to thank the Max-Planck-Instit\"ut f\"ur Extraterrestrische Physik for their invitation to the June 2024 Ringberg meeting where these ideas were presented.

\section*{Data Availability}
No new data were generated or analysed in support of this research.

\author[0000-0002-7093-7355]{A.\,Renzini}

\label{lastpage}


\vspace{1 truecm}

\begin{thebibliography}{2014}

\bibitem[\protect\citeauthoryear{Adamo et al.}
{2024}]{adamo24a}Adamo, A.,  Atek, H.,  Bagley, M.B.,  Ba–ados, E.,  Barrow, K.S.S. et al. 2024, ArXiv:2405.21054

\bibitem[\protect\citeauthoryear{Adamo et al.}
{2024}]{adamo24b}Adamo, A., Bradley, L.D., Vanzella, E.,  Claeyssens, A.,  Welch, B. et al. 2024, \nat, 632, 513

\bibitem[\protect\citeauthoryear{Akins et al.}
{2024}]{akins24}Akins, H.B., Casey, C.M., Lambrides, E., Allen, N., Andika, I.T. et al. 2014, ArXiv:2406.10341

\bibitem[\protect\citeauthoryear{Baggen et al.}
{2024}]{baggen24}Baggen, J.G.W., van Dokkum, P.,  Brammer, G.,  de Graaff, A.,  Franx, M. et al. 2024, ArXiv:2408.07745

\bibitem[\protect\citeauthoryear{Bastian \& Lardo}
{2018}]{bastian18}Bastian, N. \& Lardo, C. 2018, \araa, 56, 83

\bibitem[\protect\citeauthoryear{Bellini et al.}
{2017}]{bellini17}Bellini, A., Bianchini, P., Varri, A. L., et al. 2017, \apj, 844, 167

\bibitem[\protect\citeauthoryear{Belokurov \& Kravtsov}
{2023}]{belokurov23}Belokurov, V. \& Ktavtsov, A. 2023, \mnras, 525, 4456

\bibitem[\protect\citeauthoryear{Bianchini et al.}
{2018}]{bianchini18}Bianchini, P., van der Marel, R.P., del Pino, A., Watkins, L.L., Bellini, A. et al. 2018, \mnras, 481, 2125 

\bibitem[\protect\citeauthoryear{Bunker et al.}
{2023}]{bunker23}Bunker, A.J., Saxena, A.,  Cameron, A.J., Willott, C.J., Curtis-Lake, E. et al. 2023, \aap, 677,  A88 

\bibitem[\protect\citeauthoryear{Dekel et al.}
{2023}]{dekel23}Dekel, A., Sarkar, K.C., Birnboim, Y., Mandelker, N. \& Li, Z. 2023, \mnras, 523, 3201

\bibitem[\protect\citeauthoryear{de Mink et al.}
{2009}]{demink09}de Mink, S. E., Pols, O. R., Langer, N., \& Izzard, R. G. 2009, \aap, 507, L1

\bibitem[\protect\citeauthoryear{Elmegreen}
{2017}]{elmegreen17}Elmegreen, B.G. 2017, \apj, 836, 80

\bibitem[\protect\citeauthoryear{Farjon \& De Marchi}
{2024}]{farjon24}Farjon, K. \& De Marchi, G. 2024, \aap, 681, A20

\bibitem[\protect\citeauthoryear{Denissenkov \& Hartwick}                  
{2014}]{denissenkov14} Denissenkov, P.A.\& Hartwick, F.D.A.  2014, \mnras, 437, L21

\bibitem[\protect\citeauthoryear{Ferrara, Pallottini \& Dayal}
{2023}]{ferrara23}Ferrara, A., Pallottini, A. \& Dayal, P. 2023, \mnras, 522, 3986

\bibitem[\protect\citeauthoryear{Finkelstein et al.}
{2023}]{finkelstein23}Finkelstein, S.L., Magley, M.B., Ferguson, H.C., Wilkinf, S.M., Kartalpepe, J.S. et al. 2023, \apj, 946, L13

\bibitem[\protect\citeauthoryear{Finkelstein et al.}
{2024}]{finkelstein24}Finkelstein, S.L., Leung, G.C.K., Bagley, M.B., Dickinson, M., Ferguson, H.C. et al. 2024, \apj, 969, L2

\bibitem[\protect\citeauthoryear{Fowler}
{1966}]{fowler66}Fowler, W.A. 1966, \apj, 144, 180

\bibitem[\protect\citeauthoryear{Gratton, Carretta \& Bragaglia}
{2012}]{gratton12}Gratton, R.G., Carretta, E. \& Bragaglia, A. 2012, \aapr, 20, 50

\bibitem[\protect\citeauthoryear{H\"aberle et al. }
{2024}]{haberle24}H\"aberle, M., Neumayer, N., Seth, A., Bellini, A., Libralato, M. et al. 2024, Nature, 631, 258

\bibitem[\protect\citeauthoryear{Harikane  et al.}
{2023}]{harikane23}Harikane, Y., Ouchi, M., Oguri, M., Ono, Y., Nakajima, K. et al. 2023, \apjs, 265, 5

\bibitem[\protect\citeauthoryear{Harris  et al.}
{2006}]{harris06}Harris, W.E., Whitmore, B.C., Karakla, D. et al. 2006, \apj, 636, 90

\bibitem[\protect\citeauthoryear{Inayoshi \& Maiolino}
{2024}]{inayoshi24}Hinayoshi, K. \& Maiolino, R. 2014, ArXiv:2409.07805

\bibitem[\protect\citeauthoryear{Jeon  et al.}
{2024}]{jeon24}Jeon, J., Bromm, V., Liu, B. \& Finkelstein, S.,L. 2024, ArXiv: 2402.18773

\bibitem[\protect\citeauthoryear{Kader et al.}
{2022}]{kader22}Kader, J.A., Pilachowski, C.A., Johnson, C.I., Rich, R.M., Young, M.D. et al. 2022, \apj, 940, 76
\bibitem[\protect\citeauthoryear{Limongi \& Chieffi}
{2018}]{limongi18}Limongi, M. \& Chieffi, A. 2018 \apjs, 237, 13

\bibitem[\protect\citeauthoryear{Maiolino et al.}
{2023a}]{maiolino23a}Maiolino, R., Scholtz, J., Curtis-Lake, E., Carniani, S., Baker, W. et al. 2023, ArXiv:2308.01230

\bibitem[\protect\citeauthoryear{Maiolino et al.}
{2023b}]{maiolino23b}Maiolino, R., Risaliti, G., Signorini, M., Trefoloni, B. , Juod\v{z}balis, et al. 2023, ArXiv:2404:00504

\bibitem[\protect\citeauthoryear{Maiolino et al.}
{2024}]{maiolino24b}Maiolino, R., Scholtz, J.,  Witstok, J.,  Carniani, S.,  D'Eugenio, F. et al. 2024, Nature, 627, 59

\bibitem[\protect\citeauthoryear{Marques-Chaves et al.}
{2024}]{marques24}Marques-Chaves, R. Schaerer, D., Vanzella, E.,  Verhamme, A., Dessauges-Zavadsky, M.  et al. 2024, arXiv:2407.18804

\bibitem[\protect\citeauthoryear{Matthee et al.}
{2024}]{matthee24}Matthee, J., Naidu, R.P., Brammer, G. et al. 2024, \apj, 963, 129

\bibitem[\protect\citeauthoryear{Milone et al.}
{2017}]{milone17} Milone, A. P., Piotto, G., Renzini, A. et al. 2017, \mnras, 464, 3636

\bibitem[\protect\citeauthoryear{Neumayer et al.}
{2020}]{neumayer20}Neumayer, N.,  Seth, A. \& B\"oker, T. 2020, A\&ARv, 28, 4

\bibitem[\protect\citeauthoryear{Pacucci et al.}     
{2023}]{pacucci23}Pacucci, F., Nguyen, B., Carniani, S., Maiolino, R. \& Fan, X. 2023, \apj, 957, L3

\bibitem[\protect\citeauthoryear{Peebles}     
{1969}]{peebles69}Peebles, P.J.E. 1969, \apj, 155, 393  

\bibitem[\protect\citeauthoryear{Perez-Gonzalez et al.}     
{2024}]{pablito24}Perez-Gonzalez, P.G., Barro, G., Rieke, G.H., Lyn, J., Rieke, M. et al. 2021, \apj, 968, 4 

\bibitem[\protect\citeauthoryear{Petralia et al.}     
{2024}]{petralia24}Petralia, I.,  Minniti, D.,  Fernández-Trincado, J.G., Lane, R.R. \&  Schiavon, R.P. 2024, \aap, 688, A92

\bibitem[\protect\citeauthoryear{Piotto et al.}     
{2015}]{piotto15}Piotto, G., Milone, A. P.,  Bedin, L. R.,  Anderson, J., King, I. R. et al. 2015, \aj, 149, 91

\bibitem[\protect\citeauthoryear{Renzini, Marino \& Milone}
{2022}]{renzini22}Renzini, A., Marino, A.F. \& Milone, A.P. 2022, \mnras, 513, 2111 (Paper I) 

\bibitem[\protect\citeauthoryear{Renzini}
{2023}]{renzini23}Renzini, A., 2023, \mnras, 525, L117 (Paper II)

\bibitem[\protect\citeauthoryear{Schneider  et al.}
{2018}]{schneider18}Schneider, F.R.N. et al. 2018, \aap, 618, A73

\bibitem[\protect\citeauthoryear{Senchyna et al.}
{2024}]{senchyna24}Senchyna, P., Plat, A., Stark, D.P. \& Rudie, G.C. 2024, \apj, 966, 92 

\bibitem[\protect\citeauthoryear{Shen et al.}
{2023}]{shen23}Shen, X., Vogelsberger, M., Boyan-Kolchin, M., Tacchella, S. \& Kannan, R. 2023,  \mnras, 525, 3254

\bibitem[\protect\citeauthoryear{Smartt}
{2015 }]{smartt15}Smartt, S.J. 2015, PASA, 32, 16

\bibitem[\protect\citeauthoryear{Sukhbold et al.}
{2016}]{sukhbold16}Sukhbold, T., Erti, S., Woosley, S.E., Brown, J.M. \& Janke, H.-T. 2016, \apj, 821, 38

\bibitem[\protect\citeauthoryear{Van Dyk et al.}
{2023}]{schuyler23}Van Dyk, S.D., de Graw, A., Baer-Way, R., Zeng, W., Filippenko, A.V. et al. 2023, \mnras, 519, 471

\bibitem[\protect\citeauthoryear{Vanzella et al.}
{2023}]{vanzella23}Vanzella, E.,  Claeyssens, A., Welch, B., Adamo, A., \& Coe, D. 2023, \apj, 945, 53

\bibitem[\protect\citeauthoryear{Watanabe et al.}
{2024}]{watanabe24}Watanabe, K.,  Ouchi, M., Nakajima, K.,  Isobe, Y.,  Tominaga, N. et al.  2024, \apj, 962, 50

\bibitem[\protect\citeauthoryear{Woods, Heger \& H\"ammerle}
{2020}]{woods20}Woods, T.E., Heger, A. \& H\"ammerle, L. 2020. \mnras, 494, 2236

\bibitem[\protect\citeauthoryear{Xu et al.}
{2024}]{xu24}Xu, Y., Ouchi, M., Yajima, H., Fukushima, H., Harikane, Y. et al. 2024, ArXiv:2404.16963

\bibitem[\protect\citeauthoryear{Zhang \& Fall}
{1999}]{zhang99}Zhang Q, Fall SM. 1999, \apj, 527, L81


\end{thebibliography}
\end{document}